\documentclass[a4paper,useAMS,usenatbib,usegraphicx]{mn2e}

\usepackage{Times}

\usepackage{epsfig}

\input{psfig.sty}
\newcommand{\beq}{\begin{equation}}
\newcommand{\eeq}{\end{equation}}

\def\gtsima{$\; \buildrel > \over \sim \;$}
\def\ltsima{$\; \buildrel < \over \sim \;$}
\def\prosima{$\; \buildrel \propto \over \sim \;$}
\def\gsim{\lower.7ex\hbox{\gtsima}}
\def\lsim{\lower.7ex\hbox{\ltsima}}
\def\simgt{\lower.7ex\hbox{\gtsima}}
\def\simlt{\lower.7ex\hbox{\ltsima}}
\def\simpr{\lower.7ex\hbox{\prosima}}
\def\la{\lsim}
\def\ga{\gsim}

\newcommand{\apj}{ApJ}

\newcommand{\apjs}{ApJS}
\newcommand{\aj}{AJ}
\newcommand{\mnras}{MNRAS}
\newcommand{\aap}{A\&A}

\newcommand{\araa}{ARA\&A}

\newcommand{\pasp}{PASP}

\newdimen\hssize
\newdimen\hdsize
\usepackage{amssymb}
\usepackage{amsmath}
\usepackage{longtable}
\usepackage{lscape}
\usepackage{booktabs,caption,fixltx2e}
\usepackage[flushleft]{threeparttable}

\begin{document}
\setlength{\hbadness}{10000}

\title[PAHs in extragalactic star forming complexes]
{Polycyclic aromatic hydrocarbons in spatially resolved extragalactic star forming complexes}

\author[M. S. Khramtsova et al.]
{M. S. Khramtsova$^{1}$\thanks{E-mail: khramtsova@inasan.ru},
D. S. Wiebe$^{1}$, P. A. Boley$^{2}$, Ya. N. Pavlyuchenkov$^{1}$\\
$^{1}$Institute of Astronomy, Russian Academy of Sciences, Pyatnitskaya str. 48, Moscow 119017, Russia\\
$^{2}$Max Planck Institute for Astronomy, K\"onigstuhl 17, Heidelberg D69117, Germany}


\maketitle

\label{firstpage}

\begin{abstract}
The abundance of polycyclic aromatic hydrocarbons (PAHs) in low- and
high-metallicity galaxies has been widely discussed since the time
when detailed infrared data for extragalactic objects were first
obtained. On the scales of entire galaxies, a smaller PAH abundance
in lower-metallicity galaxies is often observed. We study this
relationship for star-forming regions in nearby galaxies,
for a sample containing more than 200 HII complexes, using
spatially-resolved observations from the Herschel Space Observatory
and Spitzer Space Telescope.  We use a model for the dust emission to
estimate the physical parameters (PAH abundance, metallicity,
ultraviolet radiation field, etc.) of these complexes. The same
correlation of PAH abundance with metallicity, as seen for entire
galaxies, is apparently preserved at smaller scales, at least when
the Kobulnicky \& Kewley metallicity calibration is used. We discuss
possible reasons for this correlation, noting that traces of
less-effective PAH formation in low-metallicity AGB stars should be
smeared out by radial mixing in galactic disks. Effective destruction
by the harder and more intensive ultraviolet field in low-metallicity environments is
qualitatively consistent with our data, as the ultraviolet field
intensity, derived from the infrared photometry, is indeed smaller in
HII complexes with lower metallicity.
\end{abstract}

\begin{keywords}
infrared: galaxies -- ISM, galaxies: photometry
\end{keywords}

\section{Introduction}

Polycyclic aromatic hydrocarbons (PAHs) are now a well-recognized
component of the interstellar medium (ISM), both in our Galaxy, and in
external galaxies. Along with very small grains (VSGs, $r \sim
100$~\AA), PAHs are believed to be the dominant source of non-stellar emission in
the near-infrared and mid-infrared ranges ($2-20~\mu$m) in star-forming galaxies. This emission
often constitutes a sizable fraction of a galaxy's total infrared
luminosity \citep{Smith07}, and cannot be explained by large grains
alone, which are too cold in the diffuse ISM to emit at these
wavelengths.  On the other hand, grains with radii much smaller than
the typically-assumed value of 0.1~$\mu$m can heat up to very high
temperatures upon a single photon absorption. Thus,
stochastically-heated VSGs can produce mid-infrared continuum
emission, while radiatively-excited bending and vibrational modes in
PAHs are the source of several spectral features in the $\sim
2-20~\mu$m wavelength range.

As an ultraviolet (UV) photon is needed to excite a PAH or VSG
particle, the emission from these particles should be related to the
overall UV emission \citep{LD01stoch,LD02}. Because of this expected connection between the
UV luminosity and the infrared (IR) luminosity, specific PAH bands, mid-IR
continuum, and their combinations with other observables are now considered as possible indicators of star
formation \citep[e.g., ][]{Calzetti07,Zhuetal2008,Kennicutt09,Treyeretal2010,Kimetal2012}. However, the UV
radiation that excites PAH molecules can also destroy these particles,
depending on the strength and hardness of the radiation
field. The harsher UV fields expected in star-forming
regions can therefore cause a significant local depletion of small
particles; thus, the relation between PAH emission and the star
formation rate (SFR) can be more complicated than a simple monotonic
function. To assess the applicability of PAH emission as a tracer of
the SFR, it is necessary to consider processes of both their formation
and destruction in detail.

Despite the significant observational progress in this field, some key
theoretical questions about the PAH lifecycle are still debated. Of
particular interest, both for PAH physics and for their usage as a SFR
indicator, is the empirically-derived relation between the metallicity
of a galaxy and the relative PAH abundance {\citep{Engelbracht05, Madden06, Draine07, Smith07, Wu07, Hunt10}}.
Specifically, there appears to exist a metallicity threshold
($12 + \log({\rm O}/{\rm H})$ from $\approx 8.0$ to 8.4 (depending on the adopted metallicity calibration) which separates galaxies into groups
with high and low PAH abundance (in
terms of the PAH fraction of the total dust mass, i.e. $q_{\rm PAH}$).

Two general scenarios has been proposed to explain the origin of this
threshold. If PAHs are synthesized in the atmospheres of
carbon-rich asymptotic giant branch (AGB) stars \citep{Latter91}, the
efficiency of their production should depend on the carbon-to-oxygen
ratio, which can be a function of metallicity. According to this
scenario, PAHs are formed less efficiently
by evolved stars in low-metallicity galaxies \citep{Galliano08}. According to the
``destructive'' scenario, PAH destruction can be more intensive in
low-metallicity environments due to the harder UV-field
\citep{Galliano05,Madden06} or stronger shocks
\citep{OHalloranetal2006}. As neither of these scenarios now seems
clearly preferable, the question of the formation and destruction of
large molecules in these environments requires additional study.

Studies of PAHs have been tightly related to the development of
infrared telescopes. Our knowledge about their presence and
importance has expanded significantly, thanks to the Infrared Space
Observatory (ISO) and especially to the Spitzer Space Telescope. The
Spitzer IRAC and MIPS passbands cover PAH spectral features, in
addition to continuum emission from larger grains. The combination of
observations at short ($3.6-24~\mu$m) and long (70 and 160~$\mu$m)
wavelengths makes it possible to determine the PAH abundance relative
to the total dust mass in extragalactic objects \citep{DL07}. However,
for many galaxies, only unresolved photometry can be performed with
Spitzer at longer wavelengths, where the spatial resolution ($\sim 38
\arcsec$) is not sufficient to measure individual fluxes from smaller
regions. It is thus not possible to resolve individual HII regions and
star-forming complexes at these wavelengths in most galaxies, except
the for Large and Small Magellanic Clouds
\citep[e.g.,][correspondingly]{Slater,Sandstrom}, M101
\citep{Gordon08}, and some others \citep[e.g.,][]{haynesetal2010}.

\cite{munmat2009} performed a spatially resolved study of the dust
properties of galaxies in the SINGS sample \citep{KenSINGS} on a scale
larger than that of individual HII complexes, using azimuthally-averaged
emission profiles. They found that the correlation between the
metallicity and PAH content is observed not only in a galaxy as a
whole, but also radially. They conclude that at large spatial scales
this correlation is more readily explained by evolutionary
effects. In that work, no distinction is made
between PAHs in star-forming regions and PAHs in the general diffuse
ISM.

We can expect that the details of PAH physics will be revealed with
more clarity in spatially-resolved observations of individual
star-forming regions and complexes, rich in bright UV-sources, both
heating PAHs and destroying them. The intensity of 8~$\mu$m emission,
and its correlation with metallicity, may depend on the source of
heating. Within (or close to) HII regions, PAHs are heated by single
ultraviolet photons produced by young stars.  However, at some
distance from these regions, the radiation field from older stellar
populations can heat PAHs as well, and their emission is thus no
longer a direct SFR tracer \citep{Boselli04,Bendo08,Crocker}. Therefore, it is
desirable to separately observe PAH emission both in the vicinity
of HII complexes, and in the general diffuse interstellar medium.

To achieve this goal, we need the ability to study individual star-forming
regions and complexes, both at the mid- and far-IR wavelengths. The resolution at 70 and
160~$\mu$m provided by the Herschel Space Observatory (about
5.2\arcsec\ and 12\arcsec, respectively) partly satisfies the requirements. 
Corresponding physical scale is about several hundred parsecs for nearby galaxies.
This resolution in most cases does not allow studying individual HII regions but
is sufficient to analyse large HII complexes and filaments. Thus, the Herschel
data successfully complement the Spitzer data at shorter wavelengths,
which have a resolution of less than 1--2\arcsec\ for IRAC the
passbands (3.6-8.0$\mu$m), and $\sim7$\arcsec\ for MIPS 24~$\mu$m.

Previously, the spatially resolved analysis of the mid-IR properties
and their relation to the metallicity was performed by \cite{Gordon08} for a
large sample of HII regions in M101. The goal of this paper is to extend this
sample significantly and to study the properties of PAH emission in
individual HII complexes within several nearby galaxies at various
metallicities, using data from the Spitzer and Herschel telescopes. We
consider possible correlations between the various derived properties
of the HII complexes, and discuss the main scenarios for PAH formation
and destruction in the context of these observations.

\section{Observational Data}

For this work, we consider 24 galaxies which were observed with both
the Spitzer Space Telescope and the Herschel Space Observatory.  All
of the Spitzer data examined in this paper were observed as part of
the SINGS survey \citep{KenSINGS}, and most of the
Herschel data were observed as part of the KINGFISH survey
\citep{KenKING}, although we also include data for 3 galaxies
(NGC~2403, NGC~6822, NGC~5194) which were observed with Herschel as
parts of other programs. In Table~\ref{table:sample}, we list the
name and morphological type of each galaxy \citep[from][]{Moustakas},
together with the proposal identification number (Prop. ID) and
principal investigator for the Herschel data presented here.
We also list the number of HII complexes, analysed in each galaxy, and
references to the works used to identify them. In the last column we
indicate the physical resolution for each galaxy probed by the Herschel instruments.
For most galaxies, the resolution is about a few hundred parsec, however, in some cases
it is about 200 pc or even less which could mean that we probe individual giant HII
regions.

For the galaxies in Table~\ref{table:sample}, we consider only data
obtained with the IRAC and PACS \citep{PACS} instruments on Spitzer and Herschel,
respectively.  We downloaded the reduced IRAC images at wavelengths of
3.6, 4.5, 5.8 and 8.0~$\mu$m from the SINGS project
website\footnote{http://sings.stsci.edu}. For the Herschel data, we
consider images at 70 and 160~$\mu$m, obtained with the PACS
photometer.  We downloaded the so-called Level 1 data from the
Herschel Science
Archive\footnote{http://herschel.esac.esa.int/Science\_Archive.shtml},
which were reduced with version 6.1.1 of the automated reduction
pipeline.  We produced final maps from the Level 1 data using version
16 of the Scanamorphos algorithm \citep{Roussel}, with a pixel scale
of 2\farcs0.

\begin{table*}
\caption{Sample of galaxies}
\label{table:sample}
\centering
\begin{tabular}{lcccccc}
\hline
Object    & Prop. ID    & Proposer       & Type     & Number of& Ref. &Physical\\
          &         &     &          & complexes  &     & resolution, kpc \\
\hline
DDO 53    & KPOT\_rkennicu\_1 & R. Kennicutt&Im        & 2        & 5   & 0.21\\
Holmberg I & KPOT\_rkennicu\_1 & R. Kennicutt&IABm      & 1        & 5  & 0.22 \\
Holmberg II & KPOT\_rkennicu\_1 & R. Kennicutt&Im        & 6        & 20 & 0.20  \\
IC 2574    & KPOT\_rkennicu\_1 &R. Kennicutt &SABm      & 7        & 5,21 & 0.23 \\
NGC 628    & KPOT\_rkennicu\_1 &R. Kennicutt&SAc       & 11       &1,2,3,4 & 0.42\\
NGC 925    & KPOT\_rkennicu\_1 &R. Kennicutt&SABd      & 23       & 2  & 0.53  \\
NGC 1097   & SDP\_rkennicu\_3 &R. Kennicutt&SBb       & 6        & 6  & 0.99  \\
NGC 2403   & KPGT\_cwilso01\_1 &C. Wilson &SABcd     & 11       & 7,8 & 0.18 \\
NGC 3184   &KPOT\_rkennicu\_1 &R. Kennicutt&SABcd     & 14       & 2,9 & 0.65 \\
NGC 3198   & KPOT\_rkennicu\_1 &R. Kennicutt&SBc       & 8        & 9  & 0.80  \\
NGC 3351   &KPOT\_rkennicu\_1 &R. Kennicutt&SBb       & 6        & 1,15 & 0.54 \\
NGC 3521   & KPOT\_rkennicu\_1 &R. Kennicutt&SABbc     & 8        & 1,9 & 0.58\\
NGC 3621   & KPOT\_rkennicu\_1 &R. Kennicutt&SAd       & 21       & 9,16 & 0.38\\
NGC 4254   &KPOT\_rkennicu\_1  &R. Kennicutt&SAc       & 10       & 3,17,18 & 0.96\\
NGC 4321   &KPOT\_rkennicu\_1 &R. Kennicutt&SABbc     & 7        & 3,18 & 0.83\\
NGC 4559   &SDP\_rkennicu\_3  &R. Kennicutt&SABcd     & 12       & 9 & 0.60\\
NGC 4725   & KPOT\_rkennicu\_1  &R. Kennicutt&SABab pec & 7        & 9 & 0.69\\
NGC 4736   &KPOT\_rkennicu\_1   &R. Kennicutt&SAab      & 7        & 1,14 & 0.30\\
NGC 5055   & KPOT\_rkennicu\_1  &R. Kennicutt&SAbc      & 5        & 3 & 0.45\\
NGC 5194   & KPGT\_cwilso01\_1 & C. Wilson & SABbc pec & 14       & 10,11 & 0.44\\
NGC 6822   & SDP\_smadde01\_3 &S. Madden &IBm       & 5        & 12,13 & 0.02\\
NGC 6946   & KPOT\_rkennicu\_1 &R. Kennicutt&SABcd     & 7        & 3,4 & 0.40\\
NGC 7331   & KPOT\_rkennicu\_1 &R. Kennicutt&SAb       & 2        & 1,14 & 0.84\\
NGC 7793   & KPOT\_rkennicu\_1 &R. Kennicutt&SAd       & 9        & 3,19 & 0.23\\ \hline
\end{tabular}
\begin{tablenotes}
 \small
      \item References: (1) \cite{Bresolin99}, (2) \cite{vanZee98}, (3) \cite{McCall85}, 
(4) \cite{Ferguson98}, (5) \cite{Croxall09}, (6) \cite{LeeSkillman04},
(7) \cite{Garnett99}, (8) \cite{Garnett97}, (9) \cite{Zaritsky}, (10) \cite{Bresolin04},
(11) \cite{Diaz91}, (12) \cite{Lee06}, (13) \cite{Hodge88}, (14) \cite{Oey93}, 
(15) \cite{Bresolin02}, {16} \cite{Ryder}, (17) \cite{Henry94}, (18) \cite{Shields91},
(19) \cite{Edmunds84}, (20) \cite{hsk}, (21) \cite{ic2574}
\end{tablenotes}
\end{table*}

\section {Methods of measurements and analysis}
\subsection{Photometry}

Since the point spread functions (PSFs) of the observations from
different instruments and passbands differ from each other, we
convolved all shorter-wavelength images to the resolution of
160~$\mu$m PACS images, using the IDL convolution procedure and
kernels provided by \cite{aniano}. Finally, all the images were
resampled to have a pixel scale of 2\farcs0.

The convolved and resampled images were used to perform aperture
photometry in the following fashion: the flux in an aperture is
measured for each HII complex for which the metallicity is available in
the work by \cite{Moustakas}. The aperture radius is chosen separately
for each HII complex, depending on its size, but they are not less 
than the size of the lowest resolution element
(12\arcsec). Therefore, no aperture corrections were
applied. The centre of an aperture is chosen geometrically.
Finally, we correct the flux for the contribution of partial
pixels.

To subtract the background emission, we estimate its average value and
standard deviation, $\sigma_{\rm ring}$, within a ring having a width
of a few arcseconds around the aperture. To exclude bright sources
(e.g., neighbouring HII complexes) from the ring, we take into account
only pixels with intensities smaller than $3\sigma_{\rm ring}$. As the
shift of an aperture by 1 pixel may significantly (by $\sim 10\%$)
change the estimated background, we move the centre of each aperture
in all directions by 1 pixel (in steps of 0.5 pixel) and calculate the
average flux for all aperture positions. This reduces the influence of
the position error, too. The standard deviation from this average is
$\sigma_{\rm pos}$. The two uncertanties, $\sigma_{\rm pos}$ and $\sigma_{\rm ring}$, 
are comparable in magnitude and in most cases contribute more or less equally to the total uncertainty. 
The final error of the aperture flux, $\sigma$, is then equal to
\begin{equation}
\sigma = \sqrt{\sigma_{\rm ring}^2 + \sigma_{\rm pos}^2}.
\end{equation}
This value varies from 1\% to 10\% for most complexes, but can be larger
for faint complexes that cannot be clearly separated from neighbouring
complexes. Locations with a signal-to-noise ratio $< 3$ at $8\,\mu$m are excluded
from the study, which means that only large and bright complexes are
analysed, as small and faint complexes are discarded because of the
convolution process. The results of photometry for all bands are presented in Table~\ref{table:Big Table} (see below).

Finally, we note that measuring reliable fluxes for low-metallicity
galaxies, like DDO~053 and Holmberg~I, is complicated by the extremely
low signal-to-noise ratio.  As a result, we had to abandon most
complexes from these galaxies, but we present fluxes for some of the
brighter complexes in Table~\ref{table:Big Table} (see below).

\subsection{Fitting HII Complex Spectra}

{\bf We apply the algorithm that has been used in \cite{Draine07}
to calculate global SEDs of SINGS galaxies, but use it to compute emission
spectra of individual HII complexes. To model a dust emission spectrum,
we adopt the \cite{DL07} approach but use an alternative method of computing
temperature distribution functions. Details are provided below.}
 
We assume that a theoretical emission spectrum of an HII complex, located at the distance $D$, is given by \citep{Draine07}
\begin{eqnarray}
f_{\nu,{\rm model}}=\Omega_{\star}B_{\nu}(T_{\star})+\frac{M_{\rm dust}}{4\pi D^2}\Bigl[(1-\gamma)p_{\nu}^{0}(q_{\rm PAH}, U_{\rm min}) \nonumber\\
+\gamma p_{\nu}^{1}(q_{\rm PAH}, U_{\rm min})\Bigr].
\end{eqnarray}
Here the first term with $T_{\star}= 5000$~K describes the stellar contribution, $M_{\rm dust}$ is the total dust mass in the HII complex and $\Omega_{\star}$ is the starlight dilution parameter.

The radiation field in a complex is described by the scaling factor
$U$, which is given in units of the interstellar radiation field $u_{\nu}^0$ in the solar
vicinity, as estimated by \cite{MRF}. Thus, the energy density of the
radiation field $u_{\nu}$ is $U \times u_{\nu}^0$. For $U$, we use the
representation given by Eq.~23 from \cite{DL07}. In this
representation, a fraction $1-\gamma$ of all the dust is exposed to
the ``minimum'' radiation field $U_{\rm min}$, while the remaining
dust is illuminated by the enhanced radiation field, having a
power-law distribution with an exponent of $\alpha$, and an upper
limit of $U_{\rm max}$. As in \cite{Draine07}, we
adopt the fixed values $U_{\rm max}=10^6$ and $\alpha=2$.
Functions $p_{\nu}^{0}$ and $p_{\nu}^{1}$ represent emission per unit frequency per
unit mass of dust (with a PAH fraction $q_{\rm PAH}$) illuminated by the minimum
radiation field and by the enhanced radiation field, respectively.

We adopt a grain model from \cite{DL07} which represents a dust mixture consisting of carbonaceous
particles (including VSGs and PAHs) and amorphous silicates. The Milky Way dust with $R_{\rm V}
= 3.1$ is utilized. For grains larger than 250~\AA\ an eqilibrium temperature is calculated for a given
starlight intensity. For smaller grains we take the stochastic heating into account and calculate
the temperature distribution functions as described by \cite{Pavyar}.

We follow the fitting procedure described in \cite{Draine07}. The
best-fit SED for an HII complex is found using the observed fluxes at
3.6, 4.5, 5.8, 8.0, 24, 70, and 160~$\mu$m. The fitting parameters are
the PAH mass fraction $q_{\rm PAH}$, the minimum starlight intensity
$U_{\rm min}$, the parameter $\gamma$ describing the fraction of dust
headed by radiation field with intensity greater than $U_{\rm min}$, the dust mass
$M_{\rm dust}$ and the solid angle $\Omega_{\star}$ of stellar
emission. The last two parameters are used for normalization only.

In the parameter grid, used for fitting, the $q_{\rm PAH}$ values range from 0.47 to 4.6\% (a step is 0.1); the minimum starlight $U_{\rm min}$
varies from 0.1 to 25 (a step is 0.1); and $\gamma$ varies from 0 to 0.3
(we sample this parameter 10 times within this range in logarithmic scale). The best-fit model is the model with
the minimum value of $\chi^2$, where $\chi^2$ is given by
\begin{equation}
\chi^2= \sum_{\rm band}{\frac{(f_{\nu,{\rm band}}^{\rm obs}-f_{\nu,{\rm band}}^{\rm model})^2}{(\sigma_{\rm band}^{\rm obs})^2+(\sigma_{\rm band}^{\rm model})^2}}.
\end{equation}
Here $f_{\nu,{\rm band}}^{\rm obs}$ and $\sigma_{\rm band}^{\rm obs}$
are the observed flux density at a given passband and the uncertainty
of this value, while $f_{\nu,{\rm band}}^{\rm model}$ and $\sigma_{\rm
  band}^{\rm model}$ are the theoretical flux density and its
``uncertainty,'' which has been proposed by \cite{Draine07} to account
for different levels of uncertainties in different bands. It is
estimated as $0.1 f_{\nu,{\rm band}}^{\rm model}$.  The best-fit
$\chi^2$ values vary from 0.5 to 3, depending on the
region. Uncertainties in the derived parameters are calculated using
the Monte-Carlo method.

\subsection{Metallicities of the HII complexes}

Many different methods have been developed to determine the
metallicity of HII regions. One of the main methods is the so-called
``direct'' $T_{\rm e}$ method, based on the electron temperature
measurement \citep[e.g.][]{Stasinska}. However, this method requires
observations of very faint nebular lines, and consequently is of
limited use.  Besides this, various other ``indirect'' methods can be
used. While numerous methods are available, metallicity estimates
obtained with different methods may differ significantly (up to 0.7
dex). Thus, regardless of which particular method is chosen, it is
important for consistency to use the \emph{same} method to evaluate
the metallicities for all HII regions.

We use the metallicities of HII complexes in the SINGS galaxies
calculated by \cite{Moustakas}. They collected spectroscopic data
from various sources and used two methods to estimate
metallicities. These methods are the theoretical relationship
presented by \cite{Kobulnicky} (hereafter, KK04) and the
empirically-derived relationship of \cite{Pilyugin} (hereafter, PT05).
In both methods, relatively strong nebular lines
({[O\,{\sc ii}]}$\lambda$ 3727; {[O\,{\sc iii}]}$\lambda\lambda$ 4959,
5007; H$\beta$) are utilized.

These methods are convenient for a number of reasons. First, estimates
obtained with them tend to bracket metallicities determined by various
other methods. In other words, the ``true'' value of metallicity
presumably lies somewhere within the range bounded by values
calculated with the KK04 and PT05 methods. Second, while other
indirect methods are recommended for objects with specific values of
luminosity, metallicity, etc., the KK04 and PT05 methods are more or
less universal.  The HII complexes in our sample possess a wide range of
physical properties, so this ``universality'' is desirable.  For
further discussions and a comparison of these methods, we refer to
the publication by \citet{Moustakas}, but some points are worth to be repeated
here.

\begin{figure}
\centerline{\psfig{figure=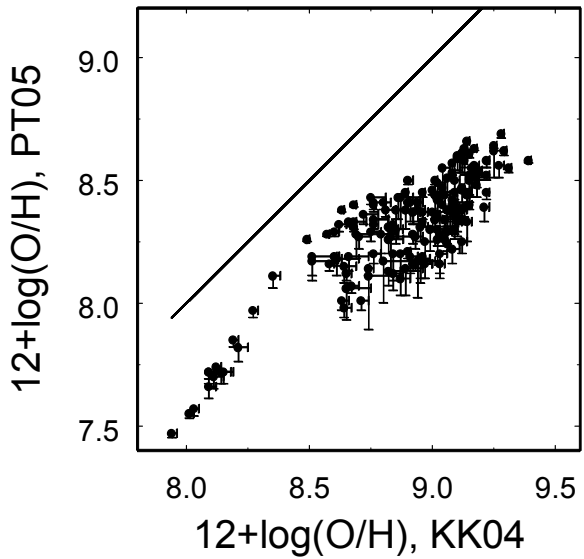,width=0.45\hdsize}}
\caption{Comparison of metallicities estimated by KK04 and PT05 methods. A straight line $y=x$ is added for clarity.}
\label{KK04vsPT05}
\end{figure}

First, metallicities determined with the KK04 method are systematically higher than
those determined with the PT05 method (Fig.~\ref{KK04vsPT05}). As the latter
metallicities are similar to
metallicities obtained with the $T_{\rm e}$ method \citep{Egorov}, the KK04 method
results are also systematically higher than ``direct'' metallicity estimates.
Second, when the PT05 metallicities are plotted versus galactocentric radius, they show
shallower gradients and much more scatter than the KK04 metallicities \citep{Moustakas}.
We see the scatter in our results, too (see below).
Furthermore, \citet{Lopez} compared these methods, and
suggested that the KK04 method produces the most reliable
metallicities for stellar populations.

\section{Results}

The results of the aperture photometry are presented in
Table~\ref{table:Big Table}, with the following columns: (1)
sequential number; (2) designation of the HII complex; (3)-(4) right
ascension and declination of the complex; (5) an aperture size; 
and (6)-(12) the fluxes at IRAC 3.6, 4.5, 5.8, 8.0, MIPS 24, PACS 70, 
and PACS 160~\micron{} bands. The results of fitting procedure are presented 
in Table \ref{table:Big Table2}, which contains the following columns: (1) sequential number; (2) designation; (3) the relative PAH
  mass abundance, $q_{\rm PAH}$; (4) $\gamma$ parameter; (5) minimum
starlight intensity $U_{\rm min}$; (6) and (7) metallicities of the
complex, determined by \cite{Moustakas} with the KK04 and PT05 methods.

\begin{figure}
\centerline{\psfig{figure=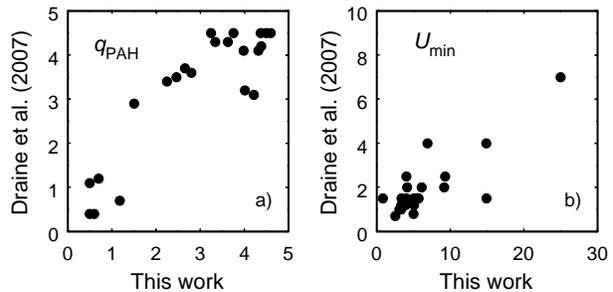,width=0.45\hdsize}}
\caption{Comparison of $q_{\rm PAH}$ (a) and $U_{\rm min}$ (b) values obtained in this work and in Draine et al. (2007).}
\label{compardraine}
\end{figure}

The values of $q_{\rm PAH}$, $\gamma$ and $U_{\rm min}$ have been
determined for each galaxy in the SINGS sample by \cite{Draine07}.  In
Fig.~\ref{compardraine}, we compare the values of $q_{\rm PAH}$ and
$U_{\rm min}$ for each galaxy, obtained by simple averaging of the
values from Table~\ref{table:Big Table2}, with the respective values
given by \cite{Draine07}. Note that in \cite{Draine07} the parameters
were obtained based on global photometry of galaxies including diffuse medium
not associated with HII complexes which can contribute a significant fraction
to the total infrared emission. The PAH abundances in both studies agree
well with each other, while our starlight intensities are
systematically higher than the values of \cite{Draine07} (the same is
true for $\gamma$ as well). This is expected, as we only consider HII
complexes, and our $U_{\rm min}$ starlight intensities are biased toward
higher values.

Since we are mostly interested in PAHs, below we consider several
relations between the metallicity and parameters related to
PAHs. Note that complexes with minimum and maximum $q_{\rm PAH}$ values
represent, correspondingly, upper and lower limits. In this study we
use the  models for grain size distributions from
\cite{WD01}, and they provide parameters for distributions only for
dust mixtures in this range of $q_{\rm PAH}$. Thus, we are unable to
measure $q_{\rm PAH}$ values lower than 0.5\% and greater than 4.6\%.
Note that $q_{\rm PAH}$ is insensitive to the band-to-band variations
which can be caused by changes in the mean ionization state or grain size
distribution.

\subsection{Derived parameters of HII complexes} 

\begin{figure*}
\centerline{\psfig{figure=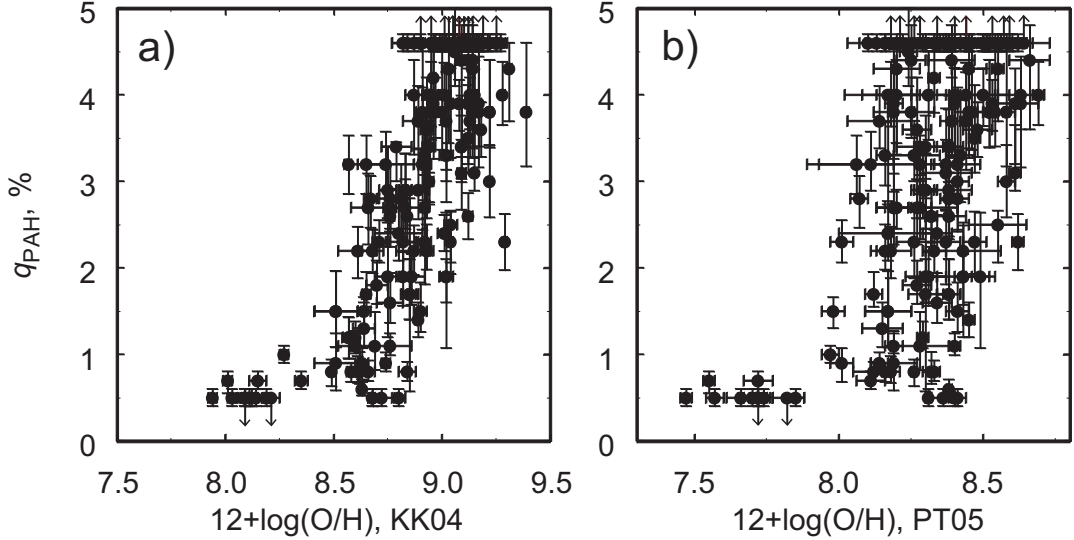,width=0.8\hdsize}}
\caption{Relation between $q_{\rm PAH}$ and metallicity for individual
HII complexes in all studied galaxies for the two metallicity calibrations.}
\label{bothmet}
\end{figure*}

We have determined mass fraction of PAHs in more than 200 HII complexes
for which metallicities are known. The combined results are presented in
Fig.~\ref{bothmet} both for the KK04 metallicity calibration and for the PT05 metallicity calibration.
The metallicity threshold inferred in previous
studies \citep{Engelbracht05,Draine07,Gordon08}, which divides HII complexes with
high and low PAH abundances, is apparently present in
Fig.~\ref{bothmet}. It is $\sim8.4$ in the KK04 scale and $\sim8.0$ in the
PT05 scale which is consistent with previous estimates. Depending on the method
of metallicity estimation some previous studies quoted different values of the threshold.
E. g., \cite{Draine07}, using the PT05 method, found the threshold to be $\sim 8.1$, while
\cite{Engelbracht05} give the threshold value of about 8.2, using empirical methods that are
close to the PT05 one. 

When the KK04 calibration is used, a tendency is seen for the abundance of PAHs
to grow with metallicity above the threshold (Fig.~\ref{bothmet}a). For the PT05 calibration there is
more scatter on the $q_{\rm PAH}$-metallicity plot that has a
staggered appearance (Fig.~\ref{bothmet}b). As this scatter would blur any correlations that
may exist in individual HII complexes, for the remainder of the present
work, we adopt the KK04 metallicities. It must be kept in mind, however, that our conclusions are
based on the specific metallicity calibration.

\begin{figure*}
\centerline{\psfig{figure=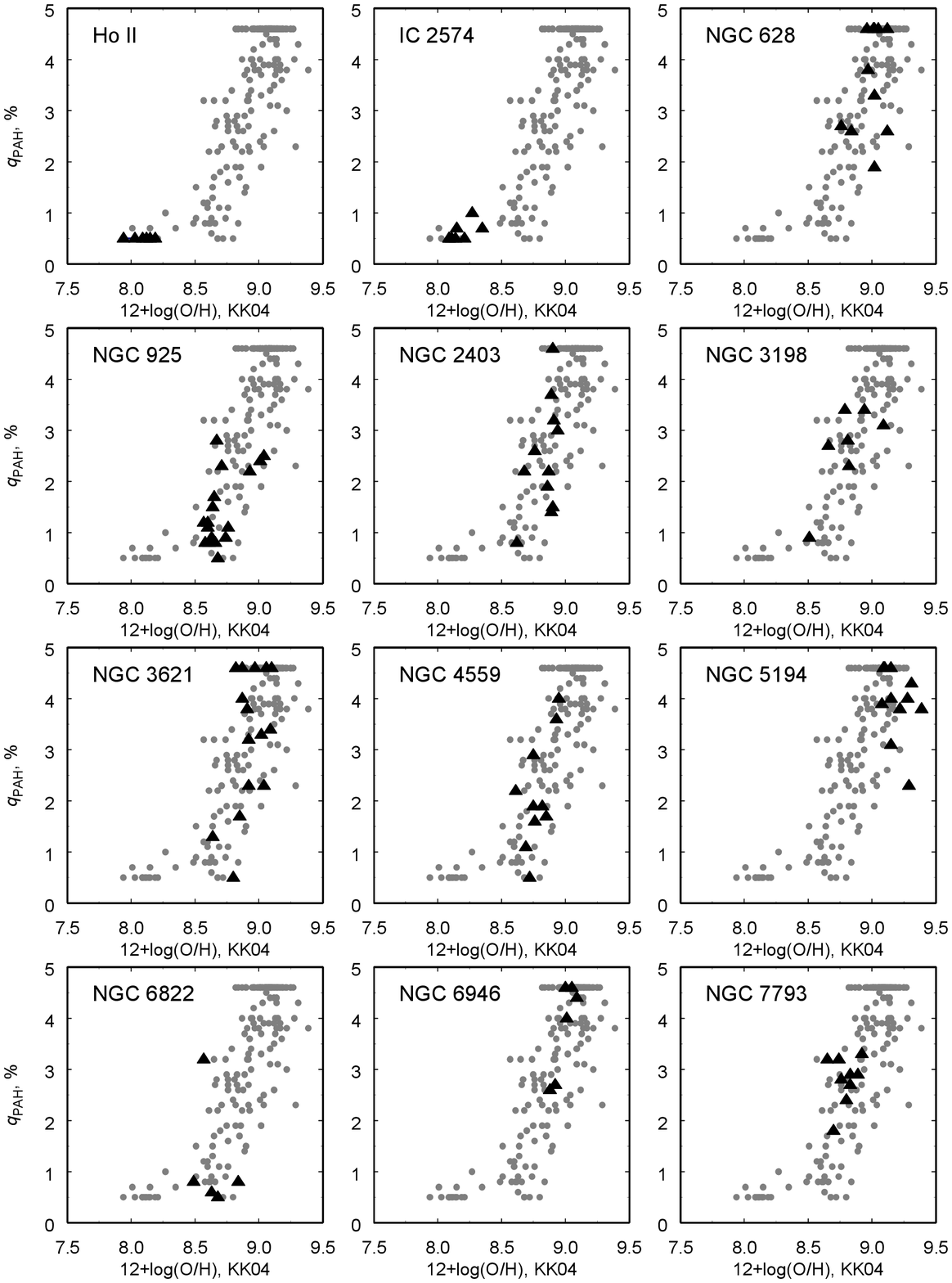,width=0.8\hdsize}}
\caption{Relation between $q_{\rm PAH}$ and metallicity for individual
HII complexes in selected galaxies. The grey points in each subplot
show the values from the entire sample, while the triangles show
only those values corresponding the galaxy indicated in the upper
left.  Galaxies which are not included in this plot have high
metallicities so that the corresponding points are concentrated in
the upper right part of the plot.}
\label{qzsep}
\end{figure*}

In Fig.~\ref{qzsep}, we show the metallicity vs. the PAH mass fraction $q_{\rm PAH}$ for
several individual galaxies.  In each subplot, we show the values
for the entire sample (over all galaxies) as grey points, and the
values for each individual galaxy (indicated by the caption) as
black triangles.

We have only a few low-metallicity HII complexes in our sample, partly
because these complexes are too faint in the infrared range.
Furthermore, the SINGS/KINGFISH sample includes only four
low-metallicity galaxies, and none of the HII complexes in these
galaxies cover the ``critical'' metallicity range from about 8.2 to
8.5 (in the KK04 scale), which apparently includes the turn-off
point. However, at higher metallicities, the tendency for the PAH
content to grow with $12+\log({\rm O}/{\rm H})$ is preserved, even
when we consider individual HII complexes instead of an entire galaxy.
This can be seen in Fig.~\ref{qzsep}, where we show data for selected
galaxies from our sample with metallicities above the ``critical''
range.  Despite the small number of data points for each galaxy, we
still believe that our data are sufficient to conclude that the
dependence of PAH content on the metallicity in star-forming
galaxies is observed both globally and {\em locally}, at the scale of
individual HII complexes. This result is consistent with the similar
conclusion made by \cite{Gordon08} for M101 galaxy.

\begin{figure*}
\centerline{\psfig{figure=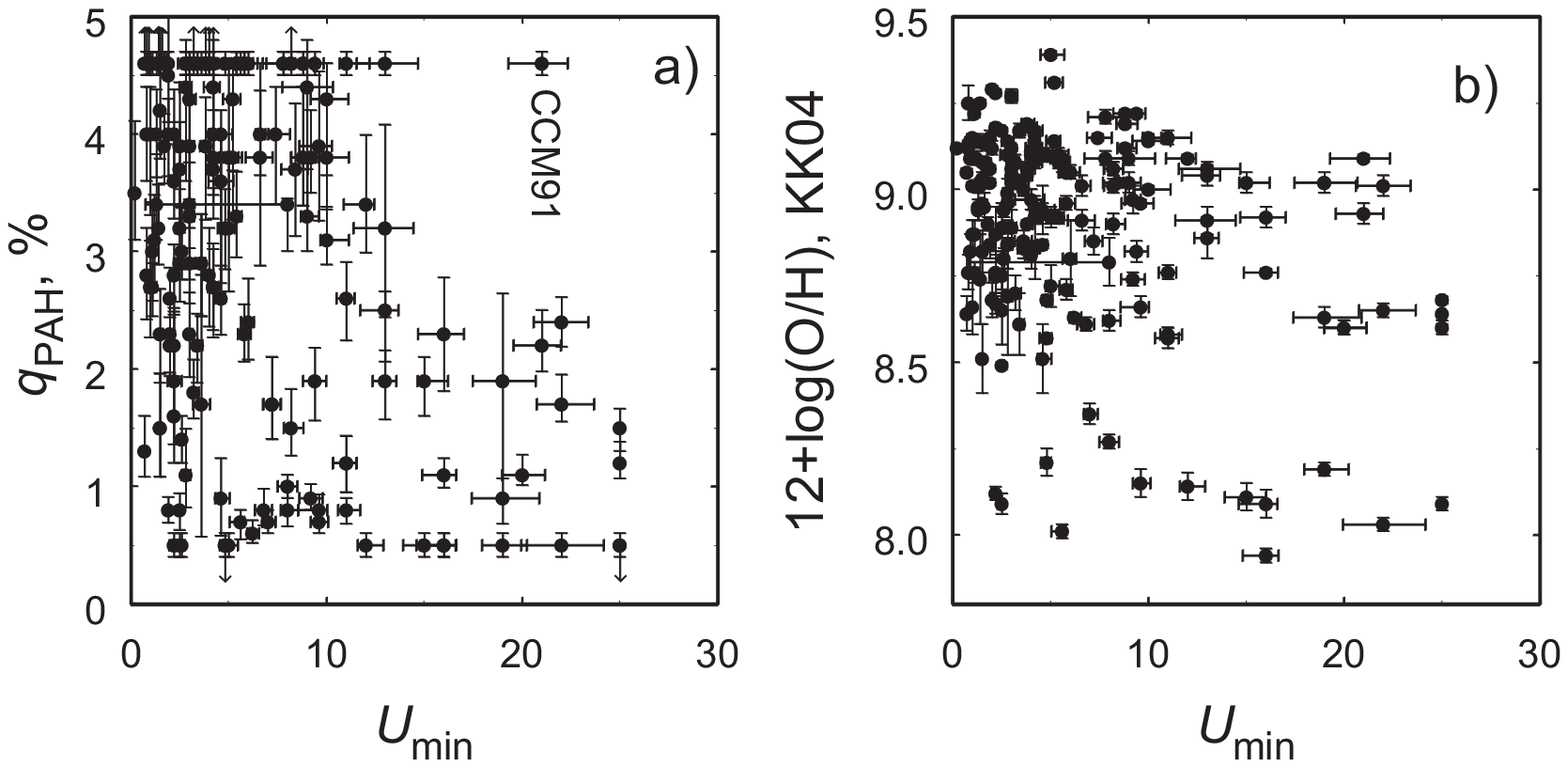,width=0.8\hdsize}}
\caption{a) Relation between $q_{\rm PAH}$ and $U_{\rm min}$ values
  for all the studied HII complexes. b) Relation between the metallicity
  and $U_{\rm min}$. Solid line shows the linear fit through all the
  data points.}
\label{uqpah}
\end{figure*}

In Fig.~\ref{uqpah}a we show how the PAH mass fraction $q_{\rm PAH}$
is related to the UV strength $U_{\rm min}$.  Solid circles correspond
to HII complexes with $q_{\rm PAH}$ less than 4.6\%, which is the
highest value we can measure with our method. Open circles are HII
complexes where the plotted $q_{\rm PAH}$ values represent lower
limits. complexes with higher $U_{\rm min}$ tend to have lower $q_{\rm
  PAH}$, albeit with significant scatter. The only noticeable
exception is the CCM91 region in M51, which has a lower limit for
$q_{\rm PAH}$ of 4.6\%, and $U_{\rm min}=21$. It is interesting that
high $U_{\rm min}$ values do indeed show some preference toward lower
metallicity HII complexes (Fig.~\ref{uqpah}b). The apparent grouping of
points in the lower part of Figure~\ref{uqpah}b simply reflects the
lack of observed complexes and galaxies in the ``critical'' metallicity
range. Probably, these intermediate systems are too faint in the
infrared to become a part of some large-scale survey, but at the same
time too metal-rich to deserve an individual study.

\subsection{Dust indices, PAH content, and metallicity}

Apart from the SED modelling, some indices can be used to characterize the dust
content. \cite{DL07} introduced the $P_{8.0}$ and $P_{24}$ indices
\begin{equation}
P_{8.0}=\frac{\nu F^{\rm ns}_{\nu}(8.0\mu {\rm m})}{\nu F_{\nu}(70\mu {\rm m})+\nu F_{\nu}(160\mu {\rm m})},
\end{equation}
\begin{equation}
P_{24}=\frac{\nu F^{\rm ns}_{\nu}(24\mu {\rm m})}{\nu F_{\nu}(70\mu {\rm m})+\nu F_{\nu}(160\mu {\rm m})},
\end{equation} 
where $F^{\rm ns}_{\nu}(8.0\mu {\rm m})$ and $F^{\rm ns}_{\nu}(24\mu {\rm m})$ are non-stellar fluxes
at 8 and 24 $\mu{\rm m}$, correspondingly. Stellar emission significantly affects
Spitzer 3.6~$\mu$m band; at 8~$\mu$m and 24~$\mu$m, the stellar
contribution is less important, but still should be taken into
account. We use the relations, given by \citet{Helou}, to
subtract stellar emission. At 70 and 160~\micron{}, the stellar contribution is negligible, and
no correction is applied to $F_{\nu}(70\mu {\rm m})$ and $F_{\nu}(160\mu {\rm m})$ fluxes.

\cite{munmat2009} proposed an analytical relation (their Eq.~A4)
between $q_{\rm PAH}$, $P_{8.0}$ and $P_{24}$, which allows estimation
of the PAH content in cases when $q_{\rm PAH}$ goes beyond the limits
of the \cite{DL07} spectra. In our data, the dependence of $q_{\rm
  PAH}$ on infrared indices look similar, but our $q_{\rm PAH}$ values
are systematically lower than those computed for the same $P_{8.0}$
and $P_{24}$ values with the analytical relation of
\cite{munmat2009}. This is, probably, again related to the fact that
we only consider HII complexes with enhanced UV radiation. In these
complexes fewer PAHs are needed to produce a certain level of 8~$\mu$m
emission compared to cases when this emission is averaged over the
galaxy as a whole (or over a significant portion of the galaxy).

\begin{figure*}
\psfig{figure=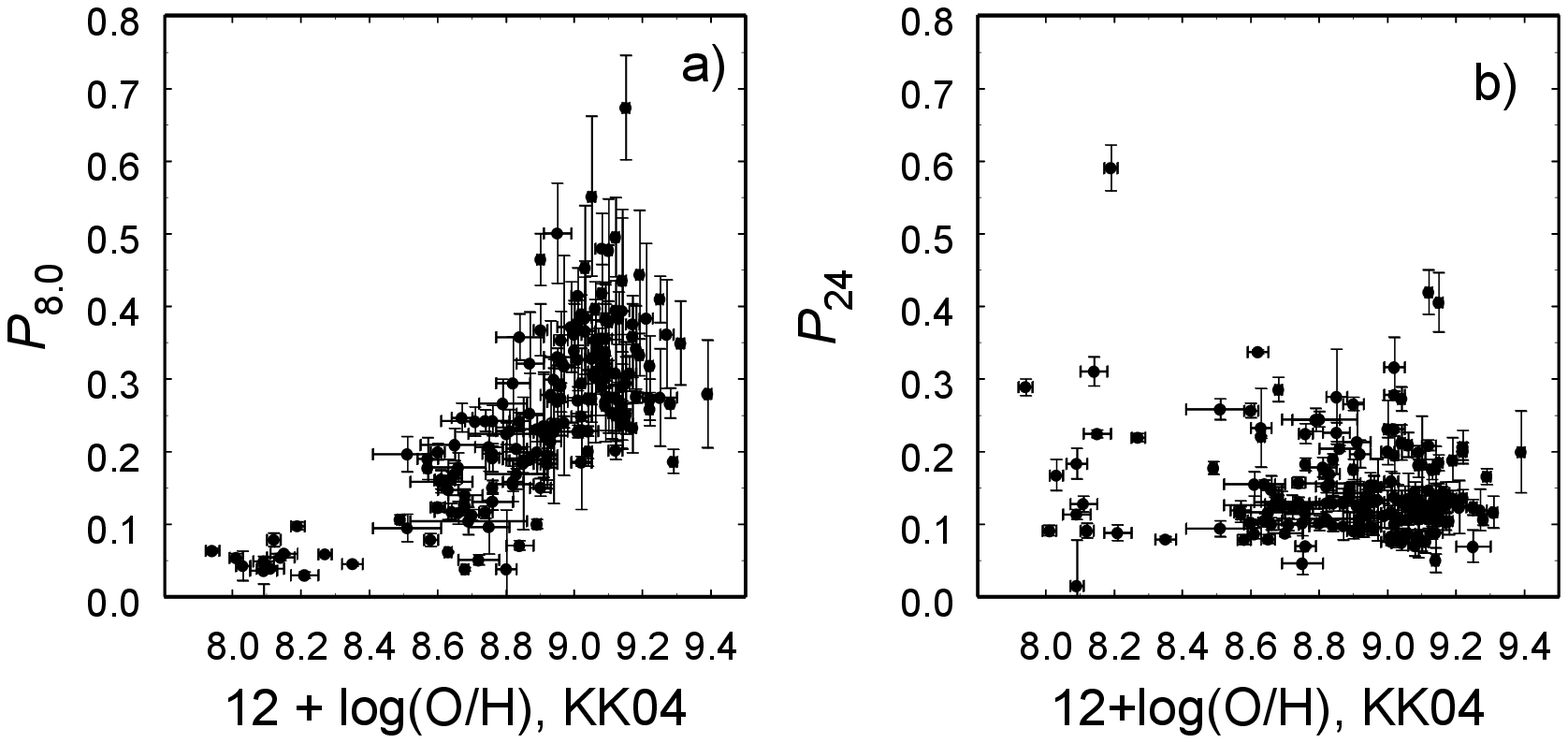,width=0.8\hdsize}
\caption{Relation between $P_{8.0}$ and $P_{24}$ indices and the metallicity of an HII complex.}
\label{P824z+}
\end{figure*}

We considered several relations between metallicity and the $P_{8.0}$
and $P_{24}$ indices. In Fig.~\ref{P824z+}, we show how $P_{8.0}$ and
$P_{24}$ are related to the metallicity of an HII complex.  $P_{8.0}$
shows nearly the same correlation with metallicity as $q_{\rm PAH}$
(Fig.~\ref{P824z+}a). On the other hand, $P_{24}$ appears to be
totally uncorrelated with metallicity (Fig.~\ref{P824z+}b).

\begin{figure*}
\psfig{figure=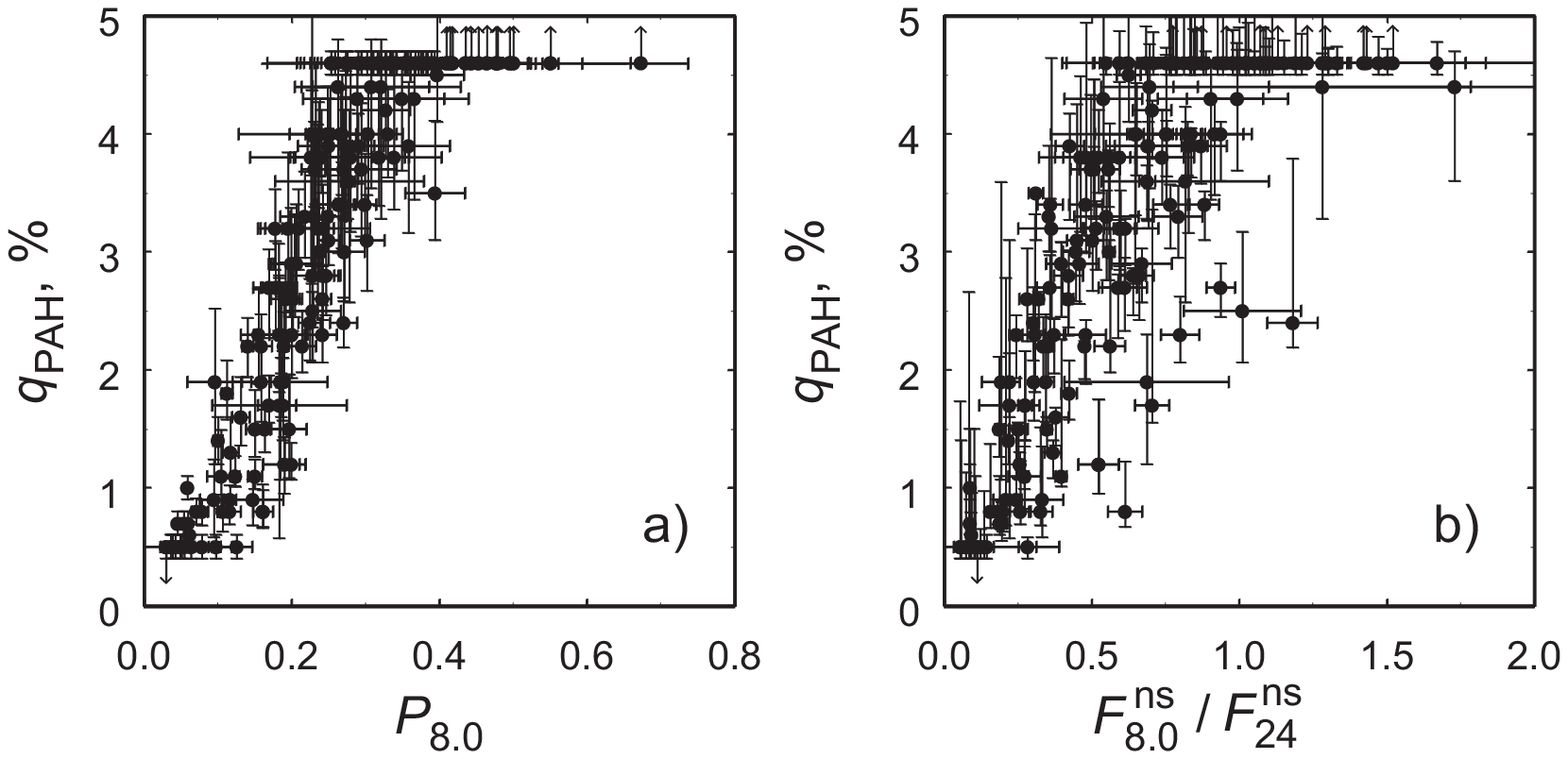,width=0.8\hdsize}
\caption{Relation between $P_{8.0}$ index, $F^{\rm ns}_{8.0}/F^{\rm ns}_{24}$ ratio, and the PAH content for individual HII complexes.}
\label{q824}
\end{figure*}

Naturally, $P_{8.0}$ is correlated with $q_{\rm PAH}$ as well (Fig.~\ref{q824}a), but this index,
unlike $q_{\rm PAH}$, is not bound to the model grid, so we used it to mark some
$q_{\rm PAH}$ values as lower and upper limits. Specifically, we estimated the
bordering values $P_{8.0}^{\rm min}$ and $P_{8.0}^{\rm max}$ that correspond to
$q_{\rm PAH}$ of 0.5\% and 4.6\%, respectively. If $P_{8.0}$ for an HII region is greater
than $P_{8.0}^{\rm max}$ or less than $P_{8.0}^{\rm min}$ we assume that the
corresponding $q_{\rm PAH}$ value is an upper or lower limit.

The absence of correlation between $P_{24}$ and metallicity implies that qualitative assessment of the PAH
content in HII complexes can be done, using the $P_{8.0}/P_{24}$ ratio or, equivalently, the $F^{\rm ns}_{8.0}/F^{\rm ns}_{24}$ ratio.
This ratio is sometimes considered to be a
quantitative measure of the PAH abundance \citep{Engelbracht05}. In
our results, it is indeed well correlated with $q_{\rm PAH}$
(Fig.~\ref{q824}b). This greatly expands the ability to quantify the
PAH evolution in HII complexes, as the $F^{\rm ns}_{8.0}/F^{\rm ns}_{24}$ ratio
does not require photometry at longer IR wavelengths.

\section{Discussion}

Our results indicate that the correlation between the PAH abundance
and metallicity may exist both globally (for entire galaxies) and
locally (for individual star-forming complexes). Here, we examine how
the local character of the $q_{\rm PAH}-({\rm O}/{\rm H})$ correlation
is related to our ability to understand the processes of PAH
formation.

According to the ``AGB'' scenario, a low $q_{\rm PAH}$ value
in low-metallicity environments reflects certain details of their
synthesis in metal-poor AGB stars. Specifically, AGB stars populate
low-metallicity galaxies as well as high-metallicity galaxies
\citep{Tosi09}, but the carbon-to-oxygen ratio in the former may be
too low for effective PAH formation.

Neither observational nor theoretical data currently seem to support
this assertion unequivocally. There are indications that C/O ratio in the interstellar
medium is
roughly constant at metallicities lower than $12+\log({\rm O}/{\rm
  H})\approx8.0$ (at the direct method scale) and grows with $({\rm O}/{\rm H})$ at higher
metallicities \citep{henry2000}. This behaviour is reminiscent of the
observed PAH behaviour, and it is tempting to assume that the
correlation between $q_{\rm PAH}$ and $({\rm O}/{\rm H})$ simply
reflects the correlation between C/O and $({\rm O}/{\rm H})$.

However, a significant number of extremely metal-poor stars with
carbon overabundance implies that dependence of the C/O ratio on
metallicity is more complicated than just ``constant, then growing.''
Also, there is still no direct observational evidence that PAHs are
formed in the atmospheres of AGB stars, even though PAHs are observed
in vicinities of some post-AGB and AGB stars with hot companions.

Circumstantial evidence is more difficult to interpret, as there are
at least two additional factors that make the situation
confusing. First, observations indicate that acetylene molecules,
believed to be precursors for PAHs, are more abundant in
low-metallicity AGB stars \citep[][and references
  therein]{Woods2012}. \cite{Woods2012} have shown with chemical
modelling that these molecules, and other hydrocarbons, form more
efficiently in the atmospheres of metal-poor stars compared to stars
of solar metallicity. Second, metal-poor stars not only produce
abundant complex hydrocarbons, but may also expel them into
circumstellar space more efficiently \citep{Mattsson}. Also, from a more
general point of view, the evolution of C and O in AGB stars is defined by many
processes that involve both synthesis and destruction, so that surface C/O ratio
(supposedly related to PAH synthesis) depends not only on the star initial metallicity and mass,
but also on time within the AGB phase \citep[e.g., ][]{lat,agbaraa}. So, it is well expected that initial metallicity would
not play a simple role in determining the PAH synthesis. At the very least, low
metallicity by itself does not guarantee fewer PAHs.

Another complication is the non-uniform and (possibly) time-dependent
metallicity distribution in most late-type galaxies. If PAHs are
synthesized in AGB stars with an efficiency that is an increasing
function of metallicity, then the PAH content should depend on the
metallicity of these parent stars (i.e., on the metallicity at
locations where these stars were formed). The original relation
between the PAH content and metallicity would be preserved if PAHs
and/or their parent stars are not able to travel in the galaxy over
significant distances (i.e., distances, over which the metallicity
changes appreciably).

The question is if AGB stars can migrate over kiloparsec scales in
their parent galaxies. Calculations show that radial mixing of stars
can flatten the metallicity gradient both in the Milky Way-type
galaxies, and in less-massive galaxies, over a time span comparable to
the lifetime (from about 100 Myr to several Gyr) of an AGB star \citep{sellwood,minchev}. The metallicity
range in the galaxies from our sample does not exceed a few tenths. To
destroy the original correlation between the PAH content and the ISM
metallicity, the star would have to have moved radially by a few
kiloparsecs prior to the AGB phase, which seems probable according to
theoretical models. If this were the case, we would expect the
metallicity gradient of evolved stars to be shallower than the
gradient of the ISM metallicity, or the gradient measured with young
objects like O stars or open clusters.

Observational evidence of this is not clear. According to some
studies, both in our galaxy and in others, the metallicity gradient
measured using planetary nebulae (PN) is indeed the same (or even
steeper) than the gradient measured with open clusters or HII regions
\citep{maciel,m81,m33,ngc6822}. According to other studies, however,
the PN metallicity gradient is shallower than that of HII regions
\citep{pn2006}. \cite{latetypestars} found that the gradient steepens
with the effective temperature of stars that have been used to measure
the gradient. This again indicates that the gradient becomes shallower
when we consider older objects, because of radial migration.

Thus, to have the PAH content correlated with the local ISM
metallicity in the ``AGB'' scenario, we need three conditions
to be satisfied. First, the efficiency of the PAH synthesis should be
related to the metallicity of the parent AGB star. Second, the parent
AGB star should stay in approximately the same metallicity environment
during its lifetime. Third, PAHs should not leave this environment due
to radial gas flows. Note that azimuthal mixing may also be important
in this respect, as at least in the Milky Way there are some azimuthal
metallicity variations \citep{azim}.

There is one more argument which may turn out to be contradictory to the
``AGB'' scenario. Supposing that VSGs are represented by PAH
clusters, it is reasonable to assume that their formation processes
are similar to those of PAHs. Thus, we may expect a correlation
between $P_{24}$ and metallicity, similar to what we see between
$P_{8.0}$ and metallicity. However, there are no signs for such
correlation (Fig.~\ref{P824z+}b).

This argument can be irrelevant if at some circumstances large grains over-power
very small grains at 24 \micron. However, most studies where distinction is made
between grains of various types indicate that in the general ISM most significant
contribution at 24 \micron\ comes from VSGs \citep{des1990,comp2011}. The same seems
to be true in PDRs as well \citep{berne1,comp08}. In the immediate vicinity of HII regions
the situation is more complicated. \cite{flagey} have shown that under certain assumption
on the grain size distributions in some regions within the Eagle Nebula emission from big
grains (BG) is indeed comparable or even exceeds emission from VSGs. \cite{Deharveng} analysed
emission maps of several HII regions at 8 and 24 \micron\ and stated that the source
of emission at 24 mum (BGs vs VSGs) cannot be indentified clearly. We have initiated our
own modelling of the infrared emission from an HII region that takes into account
grain destruction and found that the contribution from big grains at 24 \micron\ is small in comparison
to the VSG contribution (M. Kirsanova et al., submitted). Still, the question of the 24 \micron\ emission source is far from being solved.

In the ``destructive'' scenario, there is only one condition that
needs to be met. Namely, the PAH destruction efficiency should be
related to the local metallicity, i.e., through the intensity and/or
hardness of the UV field. The relation between $q_{\rm PAH}$ and the UV intensity
(characterized by $U_{\rm min}$) does seem to exist, albeit
with significant scatter (Fig.~\ref{uqpah}a). Regarding on the scatter
in the $q_{\rm PAH}-({\rm O}/{\rm H})$ and $q_{\rm PAH}-U_{\rm min}$
relation, we must note that there are two unknowns in the
``destructive'' scenario that are able to smear out metallicity
trends. These are the age of the HII region, and the initial PAH
content $q_{\rm PAH}^0$, prior to the onset of star formation in the
region.

The absence of the $P_{24}$ correlation with (O/H) (Fig.~\ref{P824z+}) does not conflict with the
``destructive'' scenario (if the emission at 24 \micron\ is indeed dominated by hot, small grains). It can be
explained by the greater stability of small grains against the
ultraviolet radiation field, compared to PAHs.

Some arguments in favour of the ``destructive'' scenario are presented in the
spatially resolved mid-IR study of the Small Magellanic Cloud (SMC) by
\cite{Bolatto} and \cite{Sandstrom}. These authors have found a large range
of the $F_{8.0}/F_{24}$ values across the SMC star-forming regions despite their
relatively uniform metallicity and concluded that the metallicity-dependent formation cannot be the
primary factor, determining the PAH abundance in the SMC. \cite{Sandstrom} also found
that the $F_{8.0}/F_{24}$ ratio correlates with $q_{\rm PAH}$, albeit with significant
scatter. In our study the correlation appears to be stronger, but we only consider star-forming
complexes, while in the SMC the greatest scatter in $F_{8.0}/F_{24}$ is apparently
observed in the diffuse medium with $q_{\rm PAH}<1.5\%$ \citep{Sandstrom}.

The factors which determine the initial PAH content in an HII region
are very diverse. First, the initial PAH molecules in a molecular
cloud may be provided by AGB stars. In this case, the possible
pre-existing $q_{\rm PAH}$--$({\rm O}/{\rm H})$ correlation,
established globally due to metallicity-dependent PAH formation, can
be further enhanced locally due to metallicity-dependent PAH
destruction. Second, even if PAHs synthesized in AGB stars are
completely destroyed as they travel through the intercloud space
($q_{\rm PAH}^0=0$), they may be synthesized in molecular clouds. As
far as we know, there are no studies that address the formation of
PAHs in molecular clouds directly. However, there are other related
studies, like \cite{bh1997}. If PAHs are indeed able to form in
molecular clouds, their observed content is determined by the balance
between the local formation (possibly metallicity-dependent) and the
local (metallicity-dependent) destruction. Third, PAHs observed in
star-forming regions may represent an intermediate stage in the
metallicity-dependent destruction of larger particles (e.g., graphitic
very small grains or amorphous hydrocarbon conglomerations).

To clarify the relative importance of all these factors, a detailed
numerical model is needed (D. Wiebe et al., in preparation). Also, it
would be useful to find a way to estimate the VSG mass fraction in HII
complexes, and to explore whether or not it shows similar correlations
as the PAH mass fraction.

\section{Summary and conclusions}

In this work we performed aperture photometry for more than 200 HII
complexes in 24 spatially-resolved nearby galaxies from the SINGS
sample. Spitzer and Herschel data are used to estimate the PAH
abundances and the UV field intensity in these HII complexes. Using
literature values of the metallicities of the HII complexes, we show
that the well-known correlation between the PAH content and
metallicity previously found for entire galaxies is preserved also for
smaller-structure components, like star-formation complexes and giant
HII regions. We lack galaxies with $12+({\rm O}/{\rm H})$ around
the threshold metallicity in our sample, so we are not able to trace the region in the
$q_{\rm PAH}-({\rm O}/{\rm H})$ parameter space where $q_{\rm PAH}$
begins to increase with metallicity.  Several arguments are presented
in favour of the ``destructive'' scenario of PAH evolution, which
relates the lower PAH mass fraction in low-metallicity environments to
a stronger UV field.  However, many factors are involved in PAHs
evolution, so more detailed modelling should be able to provide more
clear information on their evolutionary path.

\section*{Acknowledgments}
We are grateful to an anonymous referee for valuable comments and suggestions. 
This study was supported by the RFBR (grants 10-02-00231, 12-02-31452)
and by the Federal Targeted Program ``Scientific and Educational Human
Resources of Innovation-Driven Russia'' (contract with the Ministry of
Science and Education 14.V37.21.0251).  This work is based in part on
observations made with the Spitzer Space Telescope, which is operated
by the Jet Propulsion Laboratory, California Institute of Technology
under a contract with NASA, and Herschel Space Observatory, which is an ESA facility with science instruments provided by European-led Principal Investigator consortia and with important participation from NASA. We thank Hendrik Linz for explanation of
reduction process of Herschel data and essential help to this work,
and Vitaly Akimkin for useful discussion.

\onecolumn
\begin{landscape}
	\begin{longtable}{@{} l @{}cccccccccccc}
\caption{Parameters of HII complexes. Full table is available online.} \\
\label{table:Big Table}
\centering 

No.    &Object &  $\alpha$(J2000)  &  $\delta$(J2000)  &  aperture  &  $F_{3.6}$, mJy &  $F_{4.5}$, mJy & $F_{5.8}$, mJy & $F_{8.0}$, mJy & $F_{24}$, mJy          & $F_{70}$, mJy  & $F_{160}$, mJy\\ 
     &        & $^{h}\quad ^{m}\quad ^{s}$ &  $^\circ \quad ' \quad \arcsec$         & radius, $\arcsec$&           &          &          &       & & &         \\ 
\hline
\multicolumn {12}{c}{IC2574}   \\ \hline

1   &H1       & 10 28 55.60 & +68 27 54.8 & 12.0&0.01 $\pm$ 0.01& 0.01 $\pm$ 0.01& 0.07 $\pm$ 0.01& 0.15 $\pm$ 0.01 & 2.3 $\pm$ 0.1 & 3.6 $\pm$ 1.8 & 2.2 $\pm$ 2.0\\
2   &H2       & 10 28 58.91 & +68 28 27.7 & 14.0&0.12 $\pm$ 0.01& 0.10 $\pm$ 0.01& 0.21 $\pm$ 0.02& 0.31 $\pm$ 0.01 & 2.7 $\pm$ 0.1 & 45.9 $\pm$ 2.4 & 55.1 $\pm$ 6.1\\
3   &H3       & 10 28 48.50 & +68 28 02.3 & 13.0&0.34 $\pm$ 0.03& 0.36 $\pm$ 0.02& 1.10 $\pm$ 0.03& 2.4 $\pm$ 0.1 & 27.2 $\pm$ 0.3 & 259.4 $\pm$ 6.3 & 225.9 $\pm$ 8.5\\
4   &H5-6     & 10 28 50.18 & +68 28 23.3 & 14.0&0.19 $\pm$ 0.03& 0.17 $\pm$ 0.02& 0.33 $\pm$ 0.04& 0.63 $\pm$ 0.07 & 5.3 $\pm$ 0.6 & 129.7 $\pm$ 6.6 & 111.3 $\pm$ 6.8\\
5   &H8       & 10 28 43.65 & +68 28 26.4 & 12.0&0.22 $\pm$ 0.04& 0.29 $\pm$ 0.04& 0.30 $\pm$ 0.04& 0.51 $\pm$ 0.03 & 8.0 $\pm$ 0.2 & 60.7 $\pm$ 4.0 & 35.1 $\pm$ 4.7\\
6   &H10      & 10 28 39.30 & +68 28 06.9 & 12.0&0.16 $\pm$ 0.02& 0.16 $\pm$ 0.02& 0.19 $\pm$ 0.01& 0.31 $\pm$ 0.01 & 2.7 $\pm$ 0.2 & 44.5 $\pm$ 3.6 & 43.2 $\pm$ 3.4\\
7   &H13-14   & 10 28 30.84 & +68 28 08.4 & 12.0&0.06 $\pm$ 0.01& 0.04 $\pm$ 0.01& 0.08 $\pm$ 0.01& 0.09 $\pm$ 0.01 & 0.8 $\pm$ 0.1 & 8.7 $\pm$ 0.8 & 11.3 $\pm$ 2.8\\
8   & II      & 10 28 48.40 & +68 28 02.0 & 14.0&0.35 $\pm$ 0.03& 0.36 $\pm$ 0.02& 1.10 $\pm$ 0.03& 2.5 $\pm$ 0.1   & 28.0 $\pm$ 0.3 & 273.1 $\pm$ 7.2 & 236.7 $\pm$ 9.2\\
9   & III     & 10 28 50.91 & +68 25 26.1 & 14.0&0.34 $\pm$ 0.02& 0.30 $\pm$ 0.01& 0.52 $\pm$ 0.24& 0.84 $\pm$ 0.03 & 4.1 $\pm$  0.2 & 104.2 $\pm$ 2.9 &  112.1 $\pm$ 4.2 \\
\hline
\end{longtable}
\end{landscape}

\onecolumn
\begin{longtable}{@{} l @{}ccccccc}
\caption{Fitting parameters of HII complexes and metallicites. Full table is available online.} \\
\label{table:Big Table2}
\centering 
No.  &Object& $q_{\rm PAH}$ & $\gamma$,\% &$U_{\min}$&$12 + \log({\rm O}/{\rm H})$& $12 + \log({\rm O}/{\rm H})$\\ 
     &        &          &           & & KK04          & PT05 \\
\hline
\endfirsthead
\multicolumn {7}{c}{IC2574}   \\ 
\hline

1   &H1       & $0.5_{-0.1}^{+0.1}$ & $30.0_{-0.1}^{+0.1}$ & $7.6_{-0.6}^{+0.5}$ & 8.14 $\pm$ 0.04         &  7.71 $\pm$ 0.05         \\
2   &H2       & $0.5_{-0.1}^{+0.1}$ & $1.3_{-0.1}^{+0.1}$ & $16.0_{-1.4}^{+0.6}$ & 8.09 $\pm$ 0.04         &  7.72 $\pm$ 0.05         \\
3   &H3       & $0.7_{-0.1}^{+0.1}$ & $4.5_{-0.1}^{+0.1}$ & $9.6_{-0.4}^{+0.5}$ & 8.15 $\pm$ 0.04         &  7.72 $\pm$ 0.05         \\
4   &H5-6     & $0.5_{-0.1}^{+0.1}$ & $1.3_{-0.2}^{+0.1}$ & $4.8_{-0.3}^{+0.2}$ & 8.21 $\pm$ 0.04         &  7.82 $\pm$ 0.06         \\
5   &H8       & $0.5_{-0.1}^{+0.1}$ & $8.4_{-0.8}^{+0.2}$ & $12.0_{-0.4}^{+0.9}$ & 8.14 $\pm$ 0.04         &  7.72 $\pm$ 0.05        \\
6   &H10      & $0.5_{-0.1}^{+0.1}$ & $1.3_{-0.1}^{+0.0}$ & $15.0_{-1.1}^{+1.0}$ & 8.11 $\pm$ 0.04         &  7.70 $\pm$ 0.05        \\
7   &H13-14   & $0.5_{-0.1}^{+0.1}$ & $4.5_{-0.1}^{+0.1}$ & $2.5_{-0.2}^{+0.1}$ & 8.09 $\pm$ 0.03         &  7.66 $\pm$ 0.05         \\
8   & II      & $1.0_{-0.1}^{+0.1}$ & $4.5_{-0.1}^{+0.1}$ & $8.0_{-0.5}^{+0.5}$ & 8.27 $\pm$ 0.02         &  7.97 $\pm$ 0.03         \\
9   & III     & $0.7_{-0.1}^{+0.1}$ & $ 1.3_{-0.2}^{+0.1}$& $7.0_{-0.3}^{+0.4}$ & 8.35 $\pm$ 0.03         &  8.11 $\pm$0.05          \\
\hline
\end{longtable}

\label{lastpage}

\begin{thebibliography}{99}
\bibitem[Aniano et al.(2011)] {aniano} Aniano, G., Draine, B.~T., Gordon, K.~D., Sandstrom, K. 2011, \pasp, 123, 1218
\bibitem[Arsenault \& Roy(1986)]{ic2574} Arsenault, R., Roy, J.-R. 1986, \aj, 92, 567
\bibitem[Balser et al.(2011)] {azim} Balser, D. S., Rood, R. T., Bania, T. M., Anderson, L. D. 2011, \apj, 738, id. 27
\bibitem[Bendo et al.(2008)] {Bendo08} Bendo, G.~J., Draine, B.~T., Engelbracht, C.~W., et al. 2008, \mnras, 389, 629
\bibitem[Bern\`e et al.(2007)]{berne1} Bern\`e, O., Joblin, C., Deville, Y., Smith, J. D., et al. 2007, \aap, 469, 575
\bibitem[Bettens \& Herbst(1997)]{bh1997} Bettens, R. P. A., Herbst, E. 1997, \apj, 478, 585
\bibitem[Bolatto et al.(2007)]{Bolatto} Bolatto, A. D., Simon, J. D., Stanimirovi\'c, S., van Loon, J. Th. et al. 2007, \apj, 655, 212
\bibitem[Boselli et al.(2004)] {Boselli04} Boselli, A., Lequeux, J., Gavazzi, G. 2004, \aap, 428, 409
\bibitem[Bresolin et al.(1999)]{Bresolin99} Bresolin, F., Kennicutt, R.~C., Garnett, D.~R. 1999, \apj, 510, 104
\bibitem[Bresolin \& Kennicutt(2002)]{Bresolin02} Bresolin, F., Kennicutt, R.~C. 2002, \apj, 572, 838
\bibitem[Bresolin et al.(2004)]{Bresolin04} Bresolin, F., Garnett, D.~R., Kennicutt, R.~C. 2004, \apj, 615, 228
\bibitem[Calzetti et al.(2007)]{Calzetti07} Calzetti, D., Kennicutt, R.~C., Engelbracht, C.~W., Leitherer, C., et al. 2007, \apj, 666, 870
\bibitem[Compi\`egne et al.(2008)]{comp08} Compi\`egne, M., Abergel, A., Verstraete, L., Habart, E. 2008, \aap, 491, 797
\bibitem[Compi\`egne et al.(2011)]{comp2011} Compi\`egne, M., Verstraete, L., Jones, A., Bernard, J.-P., et al. 2011, \aap, 525, id.A103
\bibitem[Crocker et al.(2013)]{Crocker} Crocker, A. F., Calzetti, D., Thilker, D. A., Aniano, G., et al. 2013, \apj, 762, 79
\bibitem[Croxall et al.(2009)]{Croxall09} Croxall, K.~V., van Zee, L., Lee, H., Skillman, E.~D., Lee, J.~C., {C{\^o}t{\'e}}, S., Kennicutt, R.~C., Miller, B.~W. 2009, \apj, 705, 723
\bibitem[Deharveng et al.(2010)]{Deharveng} Deharveng, L., Schuller, F., Anderson, L.~D., Zavagno, A., Wyrowski, F., Menten, K.~M., Bronfman, L., Testi, L., Walmsley, C.~M., and Wienen, M. 2010, \aap, 523, A6
\bibitem[Desert et al.(1990)]{des1990} Desert, F.-X., Boulanger, F., Puget, J. L. 1990, \aap, 237, 215
\bibitem[Diaz et al.(1991)]{Diaz91} Diaz, A.~I., Terlevich, E., Vilchez, J.~M., Pagel, B.~E.~J., Edmunds, M.~G. 1991, \mnras, 253, 245
\bibitem[Draine et al.(2007)]{Draine07} Draine, B.~T., Dale, D. A., Bendo, G., Gordon, K. D. et al. 2007, \apj, 663, 866
\bibitem[Draine \& Li (2001)]{LD01stoch} Draine, B.~T., Li, A. 2001, \apj, 551, 807  
\bibitem[Draine \& Li(2007)]{DL07} Draine B.~T., Li A., 2007, \apj, 657, 810
\bibitem[Edmunds \& Pagel(1984)] {Edmunds84} Edmunds, M.~G., Pagel, B.~E.~J. 1984, \mnras, 211, 507
\bibitem[Egorov et al. (2012)]{Egorov} Egorov, O., Lozinskaya, T.~A., Moiseev, A.~V. 2012, arXiv:1211.3969
\bibitem[Engelbracht et al.(2005)] {Engelbracht05} Engelbracht, C.~W., Gordon, K.~D., Rieke, G.~H., Werner, M.~W., Dale, D.~A., Latter, W.~B. 2005, \apj, 628, 29
\bibitem[Ferguson et al.(1998)]{Ferguson98} Ferguson, A.~M.~N., Gallagher, J.~S., Wyse, R.~F.~G. 1998, \aj, 116, 673
\bibitem[Flagey et al.(2011)]{flagey} Flagey, N., Boulanger, F., Noriega-Crespo, A., Paladini, R., Montmerle, T., Carey, S. J., Gagn\`e, M., Shenoy, S. 2011, \aap, 531, id.A51
\bibitem[Galliano et al.(2005)]{Galliano05} Galliano, F., Madden, S. C., Jones, A. P., Wilson, C. D., Bernard, J.-P. 2005, \aap, 434, 867
\bibitem[Galliano et al.(2008)]{Galliano08} Galliano, F., Dwek, E., Chanial, P. 2008, \apj, 672, 214
\bibitem[Garnett et al.(1997)]{Garnett97} {Garnett}, D.~R., {Shields}, G.~A., Skillman, E.~D., Sagan, S. P., Dufour, R.~J. 1997, \apj, 489, 63
\bibitem[Garnett et al.(1999)]{Garnett99} Garnett, D.~R., Shields, G.~A., Peimbert, M., Torres-Peimbert, S., Skillman, E.~D., Dufour, R.~J., Terlevich, E., Terlevich, R.~J. 1999, \apj, 513, 168
\bibitem[Gordon et al.(2008)]{Gordon08} Gordon, K. D., Engelbracht, Ch. W., Rieke, G. H., Misselt, K. A., Smith, J.-D. T., Kennicutt, R. C. Jr. 2008, \apj, 682, 336
\bibitem[Haynes et al.(2010)]{haynesetal2010} Haynes, K., Cannon, J. M., Skillman, E. D., Jackson, D. C., Gehrz, R. 2010, \apj, 724, 215
\bibitem[Helou et al.(2004)]{Helou} Helou, G., Roussel, H., Appleton, P., Frayer, D. et al. 2004, \apjs, 154, 253
\bibitem[Henry et al.(1994)]{Henry94} Henry, R.~B.~C., Pagel, B.~E.~J., Chincarini, G.~L. 1994, \mnras, 266, 421
\bibitem[Henry et al.(2000)]{henry2000} Henry, R.~B.~C., Edmunds, M. G., K\"oppen, J. 2000, \apj, 541, 660
\bibitem[Hern\'andez-Mart\'inez et al.(2009)]{ngc6822} Hern\'andez-Mart\'inez, L., Pena, M., Carigi, L., Garc\'ia-Rojas, J. 2009, \aap, 505, 1027
\bibitem[Herwig(2005)]{agbaraa} Herwig, F. 2005, \araa, 43, 435
\bibitem[Hodge et al.(1988)]{Hodge88} {Hodge}, P., {Lee}, M.~G., {Kennicutt}, Jr., R.~C. 1988, \pasp, 100, 917
\bibitem[Hodge et al.(1994)]{hsk} Hodge, P., Strobel, N. V., Kennicutt, R. C. 1994, \pasp, 106, 309
\bibitem[Hunt et al.(2010)]{Hunt10} Hunt, L.~K., Thuan, T.~X., Izotov, Y.~I., Sauvage, M. 2010, \apj, 712, 164
\bibitem[Kennicutt et al.(2003)]{KenSINGS} Kennicutt, R.~C., Armus, L., Bendo, G., Calzetti, D. et al. 2003, \pasp, 115, 928
\bibitem[Kennicutt et al.(2009)]{Kennicutt09} Kennicutt, R.~C., Hao, C.-N., Calzetti, D., et al., 2009, \apj, 703, 1672
\bibitem[Kennicutt et al.(2011)]{KenKING} Kennicutt, R.~C., Calzetti, D., Aniano, G., et al. 2011, \pasp, 123, 1347
\bibitem[Kim et al.(2012)]{Kimetal2012} Kim, J. H., Im, M., Lee, H. M., Lee, M. G., et al. 2012, \apj, 760, 120
\bibitem[Kobulnicky \& Kewley(2004)]{Kobulnicky} Kobulnicky, H.~A., Kewley, L.~J. 2004, \apj, 617, 240
\bibitem[Lattanzio \& Forestini(1999)]{lat} Lattanzio, J., Forestini, M. 1999, Asymptotic giant branch stars, Proceedings of the IAU symposium 191.
Eds T. Le Bertre, A. Lebre, \& C. Waelkens. Astronomical Society of the Pacific: California, USA , 1999, p. 31
\bibitem[Latter(1991)]{Latter91} Latter, W.~B. 1991, \apj, 377, 187
\bibitem[Lee \& Skillman(2004)]{LeeSkillman04} Lee, H., Skillman, E. D. 2004, \apj, 614, 698
\bibitem[Lee et al.(2006)]{Lee06} Lee, H., Skillman, E.~D., Venn, K.~A. 2006, \apj, 642, 813
\bibitem[Li \& Draine (2002)]{LD02} Li A., Draine, B.~T. 2002, \apj, 572, 232
\bibitem[Lo\'pez-Sa\'nchez et al.(2012)]{Lopez} Lopez-Sanchez, A. R., Dopita, M. A., Kewley, L. J., Zahid, H. J., Nicholls, D. C., Scharwachter, J. 2012, MNRAS, in press, astro-ph/1203.5021
\bibitem[Maciel et al.(2005)]{maciel} Maciel, W. J., Lago, L. G., Costa, R. D. D. 2005, 433, 127
\bibitem[Madden et al.(2006)]{Madden06} Madden S.~C., Galliano, F., Jones, A. P., Sauvage, M. 2006, \aap, 446, 877
\bibitem[Magrini et al.(2009)]{m33} Magrini, L., Stanghellini, L., Villaver, E. 2009, \apj, 696, 729
\bibitem[Mathis et al.(1983)]{MRF} Mathis J.~S., Mezger, P. G., Panagia, N. 1983, \aap, 128, 212
\bibitem[Mattsson et al.(2008)]{Mattsson} Mattsson, L., Wahlin, R., H\"ofner, S., Eriksson, K. 2008, \aap, 484, L5
\bibitem[McCall et al.(1985)]{McCall85} McCall, M.~L., Rybski, P.~M., Shields, G.~A. 1985, \apjs, 57, 1
\bibitem[Minchev et al.(2011)]{minchev} Minchev, I., Famaey, B., Combes, F., Di Matteo, P., Mouhcine, M., Wozniak, H. 2011, \aap, 527, A147
\bibitem[Moustakas et al.(2010)]{Moustakas} Moustakas, J., Kennicutt, R.~C., Tremonti, C.~A., Dale, D.~A., Smith, J.~-D. T., Calzetti, D. 2010, \apjs, 190, 233
\bibitem[Mu\~noz-Mateos et al.(2009)]{munmat2009} Mu\~noz-Mateos, J. C., Gil de Paz, A., Boissier, S., Zamorano, J. et al. 2009, \apj, 701, 1965
\bibitem[Oey \& Kennicutt(1993)]{Oey93} Oey, M.~S., Kennicutt, R.~C. 1993, \apj, 411, 137
\bibitem[O'Halloran et al.(2006)]{OHalloranetal2006} O'Halloran, B., Satyapal, S., Dudik, R. P. 2006, \apj, 641, 795
\bibitem[Pavlyuchenkov et al.(2012)] {Pavyar}   Pavlyuchenkov, Ya.~N., Wiebe, D.~S., Akimkin, V.~V., Khramtsova, M.~S., Henning, Th., 2012, \mnras, 421, 2430
\bibitem[Perinotto \& Morbidelli(2006)]{pn2006} Perinotto, M., Morbidelli, L. 2006, \mnras, 372, 45
\bibitem[Pilyugin \& Thuan(2005)]{Pilyugin} Pilyugin, L.~S., Thuan, T.~X. 2005, \apj, 631, 231
\bibitem[Poglitsch et al.(2010)]{PACS} Poglitsch, A., Waelkens, C., Geis, N., Feuchtgruber, H., Vandenbussche, B., Rodriguez, L., Krause, O., et al. 2010, \aap, 218, 2P
\bibitem[Roussel(2012)]{Roussel} Roussel, H., 2012, submitted to \pasp
\bibitem[Ryder(1995)]{Ryder} Ryder, S.~D. 1995, \apj, 444, 610
\bibitem[Sandstrom et al.(2010)]{Sandstrom} Sandstrom, K.~M., Bolatto, A.~D., Draine, B.~T., Bot, C., Stanimirovi\'c S. 2010, 715, 701
\bibitem[Sellwood \& Binney(2002)]{sellwood} Sellwood, J. A., Binney, J. J. 2002, \mnras, 336, 785
\bibitem[Shields (1991)]{Shields91} Shields, G.~A. 1991, \pasp, 103, 916
\bibitem[Slater et al.(2011)]{Slater} Slater, C.~T., Oey, M.~S., Li, A., Bernard, J.-P., et al. 2011, \apj, 732, 98
\bibitem[Smith et al.(2007)] {Smith07} Smith, J.~D.~T., Draine, B.~T., Dale, D.~A., Moustakas, J. et al. 2007, \apj, 656, 770
\bibitem[Stasi\'nska(2007)]{Stasinska} Stasi\'nska, G. 2007, in Canary Islands Winter School of Astrophysics, CUP 2008, p. 1
\bibitem[Stanghellini et al.(2010)]{m81} Stanghellini, L., Magrini, L., Villaver, E., Galli, D. 2010, \aap, 521, A3
\bibitem[Tosi(2009)]{Tosi09} Tosi, M. 2009, \aap, 500, 157
\bibitem[Treyer et al.(2010)]{Treyeretal2010} Treyer, M., Schiminovich, D., Johnson, B. D., O'Dowd, M., et al. 2010, \apj, 719, 1191
\bibitem[van Zee et al.(1998)]{vanZee98} van Zee, L., Salzer, J.~J., Haynes, M.~P., O'Donoghue, A.~A., \& Balonek, T.~J. 1998, \aj, 116, 2805
\bibitem[Weingartner \& Draine(2001)]{WD01} Weingartner J.~C., Draine B.~T., 2001, \apj, 548, 296
\bibitem[Woods et al. (2012)]{Woods2012} Woods, P. M., Walsh, C., Cordiner, M. A., Kemper, F. 2012, \mnras, in press
\bibitem[Wu et al.(2007)]{Wu07} Wu, H., Zhu, Y.-N., Cao, C., Qin, B. 2007, \apj, 668, 87
\bibitem[Yu et al.(2012)]{latetypestars} Yu, J., Sellwood, J. A., Pryor, C., Chen, L., Hou, J. 2012, \apj, in press
\bibitem[Zaritsky et al.(1994)]{Zaritsky} Zaritsky, D., Kennicutt, R.~C., Huchra, J.~P. 1994, \apj, 420, 87
\bibitem[Zhu et al.(2008)]{Zhuetal2008} Zhu, Y.-N., Wu,H., Cao, Ch., Li, H.-N. 2008, \apj, 686, 155
\end{thebibliography}
\end{document}